\documentstyle[twocolumn,aps,epsf]{revtex}

\begin{document}

\title{Ripple and kink dynamics}

\author{H. Caps and N.Vandewalle}

\address{GRASP, Institut de Physique B5, Universit\'e de Li\`ege, \\
B-4000 Li\`ege, Belgium.}

\address{~\parbox{14cm}{\rm\medskip
We propose a relevant modification of the Nishimori-Ouchi model [{\em Phys.
Rev. Lett.} {\bf 71}, 197 (1993)] for granular landscape erosion.
We explicitly introduce a new parameter: the angle of repose
$\theta_r$, and a new process: avalanches. We show that the $\theta_r$
parameter leads to an asymmetry of the ripples, as observed in natural
patterns. The temporal evolution of the maximum ripple height $h_{max}$
is limited and not linear, according to recent observations. The ripple symmetry and the kink
dynamics are studied and discussed.\\ \date{Accepted for publication in Phys. Rev. E}}}

\maketitle

\hspace{-13pt} 45.70.Mg, 81.05.Rm, 64.75.+g

\narrowtext
\section{Introduction}
The formation of ripples due to the wind blowing across a sand bed
\cite{bagnold} is a complex phenomenon involving various
mechanisms such as air turbulence, non-linear fluxes of sand,
saltation, grain ejection threshold, creeping and avalanches.

Different models \cite{hoyle,nishimori,ijmpc,werner} for ripple
formation have been proposed. Some of those models consider both
hopping and rolling of grains, and they are able to reproduce
travelling ripple structures. Simulations \cite{anderson} have
also considered various additional effects like the screening of
crests, the grain reptation and the existence of a grain ejection
threshold.

The Nishimori-Ouchi (NO) model \cite{nishimori} is able to
reproduce a wide variety of different eolian structures: ripples,
star dunes, barkhanic dunes, etc... The main advantage of such
a numerical model is that all parameters can be easily tuned and
the dynamics can be deeply investigated. In a recent work \cite{caps}, we
have shown that the NO model reproduces the complex labyrinthic patterns
created in some experiments \cite{goossens}. However, the NO model leads to
unrealistic features. Among others, a cube of sand can be stable within the
NO model. It is possible to create aeolian structures with an infinite
height. In fact, the existence of a critical angle is lacking in the NO
model.

In the next section, we will propose a modification of the NO
model in order to impose a more realistic constraint to granular
surfaces. In section 3, we will study how the ripple formation depends
on the various parameters of the model. In section 4, the dynamics of the
new model will be studied and will be confronted to various situations.
Finally, a summary of our findings will be given in section 5.

\section{Models}

In this section, we present two 2-dimensional models of granular landscape evolution.
\subsection{Nishimori-Ouchi (NO)}

In the Nishimori-Ouchi model \cite{nishimori}, two modes of
granular transport are considered: {\it (i)} saltation and {\it
(ii)} reptation. The saltation process acts along the $x$ axis. It takes grains from a site
$(x-\ell,y)$ and add these grains to a site $(x,y)$. The value of the {\it
saltation length} $\ell(x,y)$ depends on the local slope of the landscape at
$(x,y)$. In this model, the saltation grains are captured by the surface when
they touch it. The granular flux $Q_{salt}$ due to the saltation process
can be written as
\begin{equation}
Q_{salt}=A\int_{x-\ell}^{x}N(x',y){\rm d}x'\ ,
\end{equation}
where $N(x,y)$ is the amount of sand per time unit and surface unit
at the position $(x,y)$. $A$ is a dimensionless constant parameter.

The NO model also considers a 2-dimensional diffusion process, i.e. the {\it reptation}.
The amount of sand displaced at $(x,y)$ is supposed to be proportional
to the local curve of the surface at that position. The flux $Q_{rept}$ due
to the reptation reads
\begin{equation}
Q_{rept}=D \nabla h,
\end{equation} with a kind of diffusion coefficient $D$. The mass
conservation imposes
\begin{equation}
\frac{\partial h}{\partial t}=-\nabla Q,
\end{equation}
where $Q=Q_{salt}+Q_{rept}$.
Therefore, the temporal evolution of the height
$h$ of sand at the position $(x,y)$ reads
\begin{eqnarray}\label{no}
\frac{\partial h(x,y,t)}{\partial
t}&=&A\left[N(x-\ell,y)\left(1-\frac{\partial\ell(x,y)}{\partial x}\right)-N(x,y)\right]\nonumber\\&+&D\Delta h.
\end{eqnarray}

As in the NO model \cite{nishimori}, we have assumed that the local slope (along the $x$ direction) mainly controls the sand flux $Q_{salt}$. Experimentally \cite{bagnold}, one observes that the wind is stronger on the crests. As a consequence, the amount of displaced sand is larger around the crests. This effect can be included in the NO model by assuming
\begin{equation}\label{flux}
\frac{\partial Q_{salt}}{\partial x}=\delta\left(1+\tanh\frac{\partial h}{\partial x}\right)\left(1 +
\varepsilon - \tanh\frac{\partial h}{\partial x}\right)
\end{equation} where the parameters $\delta$ and $\varepsilon$ give
respectively the scale factor of the saltation flux and the non-zero
asymptotic quantity of sand displaced when $\partial h/\partial x>0$. The saltation
length $\ell$ is represented by the factor $(\tanh(\frac{\partial h}{\partial x})+1)$. This
assumes that $\ell$ is maximum on the crest and that small saltation jumps
take place on the downstream part of the ripple.

\begin{figure}[H]
\begin{center}
\centerline{\epsfxsize=7.cm
\epsffile{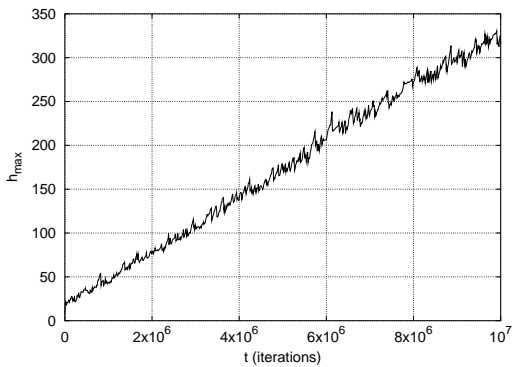}}
\vskip 0.2true cm
\caption{Evolution of the maximum height $h_{max}$ of the surface as a
function of time $t$ in the case of the NO model. The simulation parameters
are : $\delta=25$, $D=0.4$ and $\varepsilon=0.3$. The lattice size is
$101\times101$.}\label{hmaxno}
\end{center}
\end{figure}

We have numerically solved the NO model on two-dimensionnal square lattices
starting from an initially plane landscape. At each time step, a
randomly chosen site is submitted to both saltation and reptation processes
(Eqs.(4) and (5)). Figure \ref{hmaxno} shows the maximum height
$h_{max}$ reached by the profile as a function of time $t$. Time is
expressed in numerical steps (iterations).
One observes that :
{\it (i)} the growth of the maximum height $h_{max}$ is linear and {\it
(ii)} there is no saturation. This means that ripples become infinitely
high! This unrealistic behavior comes from the right hand side of
Eq.(\ref{no}). Indeed, when the first term becomes larger, the sand deposit
becomes more important than the relaxation. As time goes
on, a net growth of the profile is thus recorded. This situation
will remain unchanged since the ratio between deposit and
relaxation is fixed once by the values of $\delta$, $\varepsilon$ and $D$.
Consequently, $h_{max}$ grows {\it ad infinitum}. One should note that this
pathology implies the possibility of making cubes of sand since nothing
prohibits vertical walls. In order to get a more
realistic situation, an additional term should be included in the
NO model.

\subsection{Saltation-Creep-Avalanche (SCA)}

Granular materials are characterized by the {\it angle
of repose $\theta_r$} \cite{duran}. If the slope of a granular
pile exceeds $\tan(\theta_r)$, an avalanche is created and
stabilizes the surface. The angle of repose depends on the
grain characteristics: size distribution, shape, rugosity, and density.
Typically, this angle $\theta_r$ is about $37^\circ$ for sand \cite{bagnold,duran}.
Since it is an intrinsic property of the
granular material, the relevant parameter $\theta_r$ should be considered in
numerical models.

\begin{figure}[H]
\begin{center}
\centerline{\epsfxsize=7.cm
\epsffile{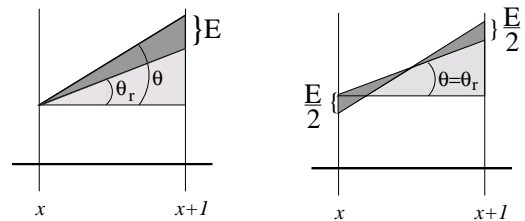}}
\vskip 0.2true cm
\caption{Avalanches lead to a profile stabilisation: {\it (left)}
the local slope $\tan(\theta)$ is larger than the critical value
$\tan(\theta_r)$, the surface is unstable.
{\it (right)} The avalanche brings the local slope to a new stable
value $\tan(\theta_r)$. Note that $E=\left|\nabla h\right|-\tan(\theta_r)$.}\label{thetar}
\end{center}
\end{figure}

Let us consider a square lattice with discrete spatial variables
$x$ and $y$. The stability condition with respect to the angle of repose
$\theta_r$ reads
\begin{eqnarray}\label{stab}
\left|h_x-h_{x+1}\right|&<&\tan(\theta_r)\nonumber\\\left|h_y-h_{y+1}\right|&<&\tan(\theta_r)
\end{eqnarray}
where the left hand side is the discrete counterpart of the
spatial derivative of the height of sand $h$ at the position $(x,y)$.
If Eq.(\ref{stab}) is not verified for a specific direction ($x$ or $y$), an avalanche occurs as
illustrated in Figure \ref{thetar}. In this case, the excess of
sand $E=\tan(\theta)-\tan(\theta_r)$ (with $\tan(\theta)=\frac{\partial
h}{\partial i}$, with $i=x$ or $i=y$ depdending on which condition (\ref{stab}) is not verified) is supposed to be equitably distributed between $x$ and $x+1$ if the avalnche occure in that direction. The same applies for the $y$ direction. Taking those considerations into account,
Eq.(\ref{no}) becomes
\begin{eqnarray}\label{nomod}
\frac{\partial h(x,y,t)}{\partial t}&=&A\left[N(x-\ell,y)\left(1-\frac{\partial\ell(x,y)}{\partial x}\right)-N(x,y)\right]\nonumber\\&+&\Theta(E)\ \frac{E}{2}+D\ \Delta h,
\end{eqnarray}
where $\Theta$ is the Heavyside step function and
\begin{equation}
E=\left|\nabla h\right|-\tan(\theta_r).
\end{equation}
The new model is called the Saltation-Creep-Avalanche (SCA) model. It will
be investigated in the present paper and numerical results will be compared with respect
to NO landscapes. One should note that we have numerically solved
Eq.(\ref{nomod}) using the same flux law (\ref{flux}) as in the original NO
model.

\section{Ripple formation}

In this section, we investigate the influence of the four parameters
$\delta$, $\varepsilon$, $D$ and $\theta_r$ on the formation
of the ripples in the SCA model. All the simulations presented herebelow
are made on $150\times 150$ square lattices. The evolution of the landscape
is typically stopped after $10^7$ iterations. The default parameter values are
$\delta=12.5$, $\varepsilon=0.3$, $D=0.4$ and $\theta_r=37^\circ$.

\subsection{Flux parameter $\delta$}

The saltation process obeys Eq.(\ref{flux}) in which the
parameter $\delta$ controls the flux $Q_{salt}$. A small arbitrary value
of $\delta$ implies that small amounts of sand are displaced along small
saltation lengths, while larger values of $\delta$ lead to a more
important transport of sand over long distances. Experimentally
\cite{bagnold}, one observes that the ripple wavelength (the distance
between two consecutive crests) grows with the saltation length $\ell$.

We have performed simulations varying $\delta$ from 1 to 100. For
very small values of the parameter ($\delta<7$), no ripple appears
and the surface remains irregular. This threshold is not encountered
in the NO model and is certainly due to the angle of repose. For small
values of the flux parameter, not enough sand is displaced and the surface
is continuously smoothed by avalanches.

Above this threshold $\delta \simeq 7$, a set of several ripples appears.
The density of ripples depends on $\delta$. The larger $\delta$ is, the
smaller is the density of ripples. For example, after $5\ 10^6$ iterations,
the density of ripples is 0.04 for $\delta=10$, while there is only a single
ripple (density of 0.006) if $\delta=85$. The SCA model is thus in good
agreement with observations: the wavelength of the ripple pattern grows
with the saltation length $\ell$. In the case of the NO model, the density of ripples
grows non-linearly with $\delta$ in opposition with experimental
observations.

\subsection{Asymmetry parameter $\varepsilon$}

\begin{figure}[H]
\begin{center}
\centerline{\epsfxsize=7.cm
\epsffile{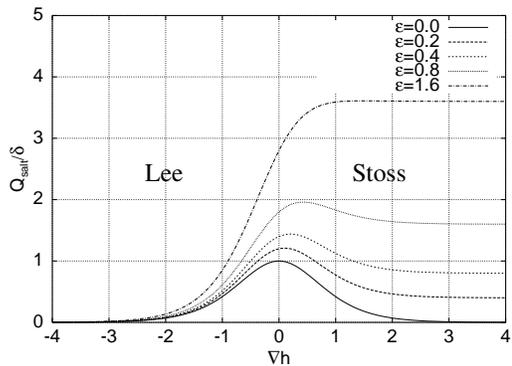}}
\vskip 0.2true cm
\caption{The normalized saltation flux $Q_{salt} / \delta$ of
Eq.(\ref{flux}) as a function of the spatial derivative $\nabla h$ of the
height of sand $h$. Different values of the parameter $\varepsilon$ are
illustrated.}\label{epsilon}
\end{center}
\end{figure}

In the SCA model, we have assumed a non-zero value for the flux of
sand $Q_{salt}$ displaced on the stoss slopes (which are upstream) of the
ripples (see Eq.(\ref{flux})). This hypothesis is supported by
numerical results \cite{nishimori}. If the parameter $\varepsilon$ is non
zero, the saltation flux becomes asymmetric (see Figure \ref{epsilon}).
Sand is always taken from the stoss slope, while $Q_{salt}$ tends to zero
on the lee slope. This comes from the fact that sand on the lee slope is
confined in wind vortices.

Practically, we have considered values of $\varepsilon$ ranging
from 0 to 2. Our observations are: {\it (i)} no ripple appears if
$\varepsilon=0$. {\it (ii)} for small values of $\varepsilon$
(typically $0<\varepsilon<0.2$), ondulations appear in the landscape. {\it
(iii)} As $\varepsilon$ becomes larger, the number of ripples present in the
system grows (from $5$ at
$\varepsilon=0.5$ to $11$ for $\varepsilon=1.8$). Moreover, ripples
are more asymmetric (windward slopes smaller than downstream slopes)
for large values of $\varepsilon$. We will study this asymmetry
in section 4.

In the case of the NO model we observe that: {\it (i)} if $\varepsilon=0$
very small ripples appear. {\it (ii)} when the parameter reaches a
threshold $\varepsilon \simeq 0.2$ larger ripples appear. Note that the
wavelength of the ripples does not depend on $\varepsilon$ above that
threshold. The fact that ripples appear for $\varepsilon=0$ in the NO model and do not appear within the SCA model is due to the existence of an angle of repose. Indeed, avalanches tend to smooth the landscape.

In summary, the flux $Q_{salt}$ should be asymmetric in order to create ripples. For
positive slopes, a minimum value of this flux is required.

\subsection{Reptation coefficient $D$}

The parameter $D$ plays the role of a diffusion coefficient for
smoothing the irregularities along the surface. We have considered values
of $D$ ranging from 0 to 1. Once again, no ripple appears for very small values of
that parameter. Ripple appearance is allowed in the interval $0.1\leq D\leq
0.5$. If $D<0.1$ the surface remains irregular. There is nearly no
interaction between the cells because of the small diffusion coefficient.
Note that the same behavior is observed in the NO model. As $D$ becomes larger, the ripple wavelength increases. Finally, for
$D>0.5$ the diffusion coefficent becomes too large and the surface remains
smooth and flat. For the NO model, no ripple appears when $D>1.8$, a larger
value than in the case of the SCA model.

\subsection{Angle of repose $\theta_r$}

Parameters $D$ and $\theta_r$ are intrinsic parameters of the granular
material, while $\delta$ and $\varepsilon$ should be related to the wind.
At each iteration, the reptation acts and moves grains. On the other hand,
an avalanche occurs if and only if the local slope of the surface exceeds
the angle of repose. The role played by both $D$ and $\theta_r$ parameters
is thus different: $D$ is a diffusion coefficient, which controls the flow
of sand between neighboring cells, while the angle of repose $\theta_r$
governs the local slope of the granular surface.

We considered typical values of $\theta_r$ between $20^\circ$ and
$40^\circ$. For ripples of the same wavelength, we have noted that small
angles of repose imply small ripple heights. The angle of repose is thus responsible of the value of the ratio $h_{max}/\lambda$, where $\lambda$ is the ripple wavelength. Indeed, a slope larger than $\tan(\theta_r)$ is reduced to a value $\tan(\theta_r)$ by avalanches. We have also noted the appearance of ripples for all tested values of $\theta_r$.

\section{Ripple dynamics}

In this section, we investigate the dynamics of the SCA model. Three
different aspects will be considered: the ripple height, the ripple shape
and the formation of kinks. Again, the default parameters values are 
$\delta=12.5$, $\varepsilon=0.3$, $D=0.4$ and $\theta_r=37^\circ$.

\subsection{Ripple amplitude $h_{max}$}

\begin{figure}[H]
\begin{center}
\centerline{\epsfxsize=7.cm
\epsffile{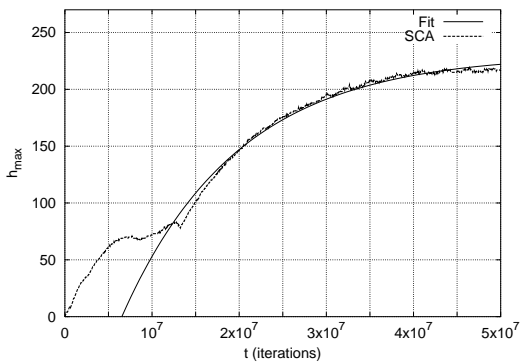}}
\vskip 0.2true cm
\caption{Evolution of the maximum height $h_{max}$ of the landscape as a
function of time $t$ in the case of the SCA model. The data have been fitted using Eq.(\ref{fit}). The fit parameters are $a=230.8\pm 0.6$, $b=377.3\pm 3$ and $c=1.33\ 10^7\pm 0.01\ 10^7$. The simulation parameters are the sames as those of Figure \ref{hmaxno}. The angle of
repose is $\theta_r=37^\circ$. When all ripples have merged, the system
contains only one ripple for which the height is limited by the angle of
repose.}\label{hmaxnomod}
\end{center}
\end{figure}

Figure \ref{hmaxnomod} presents the temporal evolution of the maximum ripple
height $h_{max}$ in the case of the SCA model. One should note the
saturation of $h_{max}$ for long times. This figure should be compared with Figure \ref{hmaxno}.

\begin{figure}[H]
\begin{center}
\centerline{\epsfxsize=8.5cm \epsfysize=9.cm
\epsffile{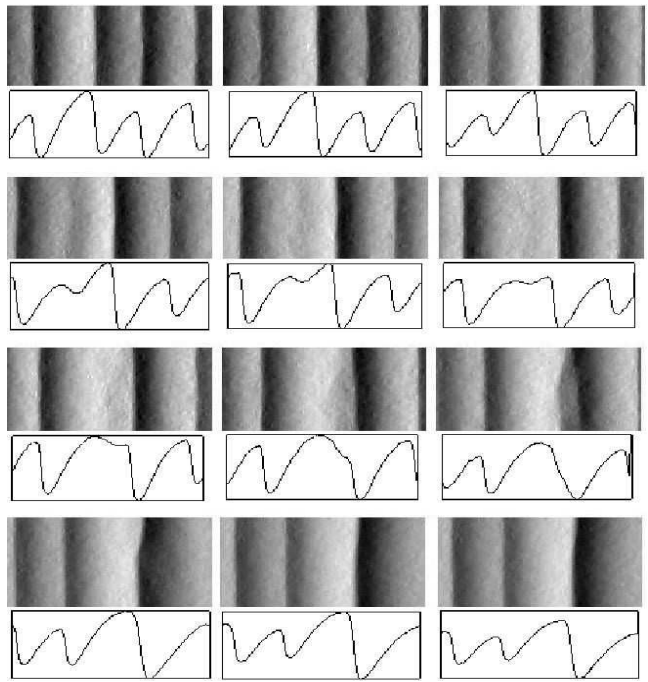}}
\vskip 0.2true cm
\caption{The merge of two ripples. The wind blows from left to right. The
profile of the landscape is shown below each picture. {\it (top row)} the
smallest ripple travels faster than the largest one. {\it (second row)} The
small ripple climbs on the large one. {\it (third row)} An avalanche is
initiated by the arrival of the small ripple on the crest. {\it (bottom
row)} The small ripple `feeds' the large one. Note that the profiles have been rescaled in order to be compared.}\label{jump}
\end{center}
\end{figure}

In Figure \ref{hmaxnomod}, many gaps can be observed in the early stages of evolution.
Those gaps come from the absorption of small ripples by the largest ones.
Since a small ripple travels faster than a large one \cite{bagnold}, the
large ripple tends to absorb the small. Figure \ref{jump}
presents the evolution of a part of the lattice for different stages of the
simulation. For each picture, a transverse view of the landscape is shown.
One should note that the merge occurs as follows: first, the small ripple
climbs on the larger one. The arrival of the small ripple on the crest
causes an avalanche. A part of the small ripple falls in
the avalanche, while the remainder feeds the larger one. It results in a
fast growth of the largest ripple even it moves slowly.

Asymptotically, only one ripple occupies the whole lattice, and the time needed for a global merge becomes larger as $\delta$ decreases. If $\delta$ is large, the initial density of ripples is small. In this case, the global merge is obtained after a small number of collapses. In opposition, a large number of merges is necessary for small values of $\delta$. If one assumes that ripple collapses is a size independent process, one can understand that the global merge is faster when $\delta\gg 1$.

The main characteristics of $h_{max}(t)$ are {\it (i)} gaps in the early stages of ripple formation, and {\it (ii)} a saturation for long times. Looking for details in Figure \ref{hmaxnomod}, one could see that gaps are allways followed by the same kind of growth. Indeed, the maximum ripple height evolves according to an exponential growth law
\begin{equation}\label{fit}
h_{max}=a-b\,\exp\left(\frac{-t}{c}\right),
\end{equation}
where $a$, $b$ and $c$ are fitting parameters. This law is shown in Figure \ref{hmaxnomod}. One should see that the fit has been performed after the primary gaps, e.g. for $t>10^7$ iterations. 

In summary, we have seen that, contrary to the NO model, the SCA model naturally leads to a non-linear evolution of $h_{max}$. Note that this kind of behavior has been experimentally observed in \cite{frette,stegner}. Moreover, the SCA model predicts a saturation of $h_{max}$ after a finite time, as expected and observed. The saturation value depends essentially on $\theta_r$. The larger is $\theta_r$, the higher are the ripples.

\subsection{Ripple aspect ratio $\sigma$}

\begin{figure}[H]
\begin{center}
\centerline{\epsfxsize=7.cm
\epsffile{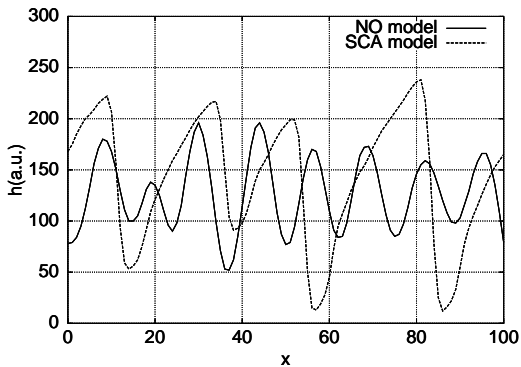}}
\vskip 0.2true cm
\caption{Height $h$ of sand (in arbitrary units) as a function of the spatial coordinate $x$.
Both NO and SCA models are shown. The curves have been rescaled in order to
be compared.}\label{sym}
\end{center}
\end{figure}

Under a non-oscillating wind, a ripple has generally an asymmetric
shape \cite{bagnold,goossens}. The stoss slope is indeed smaller
than the lee slope. In Figure \ref{sym}, the profile of the ripples is
shown for both NO and SCA models. Ripples are more symmetric in the NO
model than in the SCA model.

In order to measure the aspect ratio $\sigma$ of a ripple, we have
measured the ratio of the horizontal projections of both sides of the ripples. Let us
call $x_s$ the projection of the stoss slope on the horizontal $x$ axis, and
$x_l$ the projection of the lee slope. The apect ratio of a ripple is
defined as
\begin{equation}
\sigma=\frac{x_s}{x_l}.
\end{equation}
The measure of $\sigma$ is averaged over all the ripples of the landscape. If
$\sigma =1$ the ripples are symmetric, while $\sigma\neq 1$ implies an
asymmetry.

\begin{figure}[H]
\begin{center}
\centerline{\epsfxsize=7.cm
\epsffile{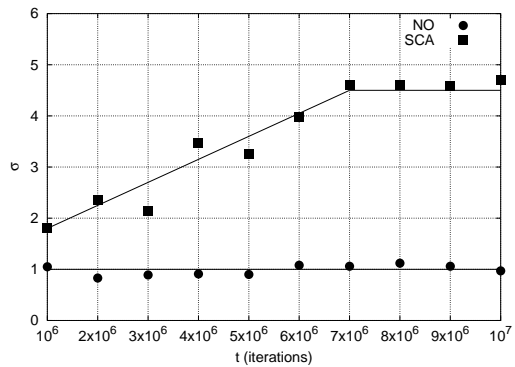}}
\vskip 0.2true cm
\caption{Temporal evolution of the aspect ratio $\sigma$. Both NO and SCA
models are illustrated. The parameter values are $\delta=12$, $\varepsilon=0.3$, $D=0.4$ and
$\theta_r=37^\circ$.}\label{s_no_sca}
\end{center}
\end{figure}

In Figure \ref{s_no_sca}, the temporal evolution of $\sigma$ for both NO and SCA models are reported. Typical values of $\sigma$ range between 1 and 5 in the SCA model. In the NO model, $\sigma$ remains close to 1.

In the case of the SCA model, one should note that $\sigma$ grows linearly with time, and then remains close to a saturation value. The appearence of a saturation seems coherent since only one ripple occupies the whole lattice after a finite time. The angle of repose limits the height of that ripple and its morphology. 

In order to compare the four parameter dependencies on $\sigma$, we have considered a reference situation which corresponds to the default parameters. Varying the value of a parameter while keeping constant the others allows us to define the influence of that parameter on the temporal evolution of the aspect ratio. Herebelow, we describe the results obtained for each of the four parameters $\delta$, $\varepsilon$, $D$ and $\theta_r$. One should note that in all cases, the temporal evolution of $\sigma$ begins with a quasi-linear law, followed by a saturation.

The aspect ratio increases with the parameter $\delta$. Since the ripple wavelength directly depends on $\delta$, that result means that small ripples are less asymmetric than large ones. This result is in good agreement with observations \cite{bagnold,goossens}. 

%Starting from a value $\sigma=1.5$ at $t=10^6$ iterations, the aspect ratio becomes $\sigma=2.6$ after $10^7$ iterations for $\delta=8$. If $\delta=25$, $\sigma=2.5$ at the beginning and grows to the value $\sigma=5$ after $10^7$ iterations. 

\begin{figure}[H]
\begin{center}
\centerline{\epsfxsize=7.cm
\epsffile{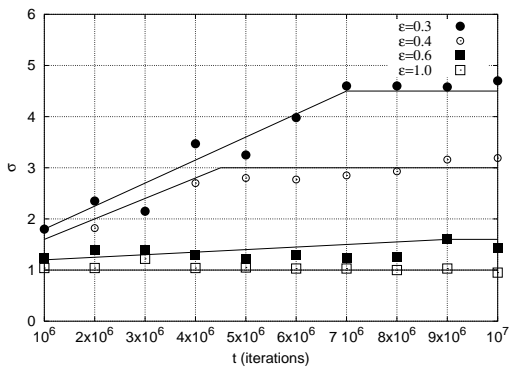}}
\vskip 0.2true cm
\caption{Temporal evolution of the aspect ratio $\sigma$ in the case of the SCA model. Different values of $\varepsilon$ are illustrated.}\label{s_sca_e}
\end{center}
\end{figure}

Figure \ref{s_sca_e} presents the temporal evolution of the aspect ratio $\sigma$ for different values of $\varepsilon$. One can see that small ripples (for small $\varepsilon$ values) are less asymmetric. Indeed, $\sigma$ is larger for $\varepsilon=1$ than for $\varepsilon=0.3$. We have seen in Section III.B. that the ripple wavelength becomes larger as $\varepsilon$ increases. One should also note that $\sigma$ saturates more rapidly when $\varepsilon$ is small. This is consistent with the fact that the time needed for a global merge is shorter in the case of large ripples.

As observed for $\delta$ and $\varepsilon$, the aspect ratio follows any variation of $D$. This results is in agreement with those observed previously in this section. Indeed, we have seen in Section III.D. that the ripple wavelength increases as $D$ becomes larger. 

We have taken realistic values of $\theta_r$ between 30${}^\circ$ and 40${}^\circ$. The aspect ratio $\sigma$ is found to be independant of the value of $\theta_r$. This result is outside the scope of this paper and will be studied in the future.

\subsection{Kink dynamics}

Although the  wind blows in a single direction, the resulting ripples are not strictly perpendicular to the wind direction. There are defects. Two different classes of defects can be found: {\it kinks} which are fusion points of two crests, and {\it antikinks} which are endings of crests. In Ref.\cite{caps}, we have studied in the NO model how defects modify a landscape morphology when a brutal change occurs in the wind direction. We have also shown in Ref.\cite{caps} that under a constant wind direction, the density of defects decreases as time goes on. In the present section, we will focus on the processes leading to this decrease in the SCA model. 

After extensive simulations, we conclude that some kinks can propagate during very long times, while others disappear quite rapidly.

\begin{figure}[H]
\begin{center}
\centerline{\epsfxsize=9.cm \epsfysize=6.cm
\epsffile{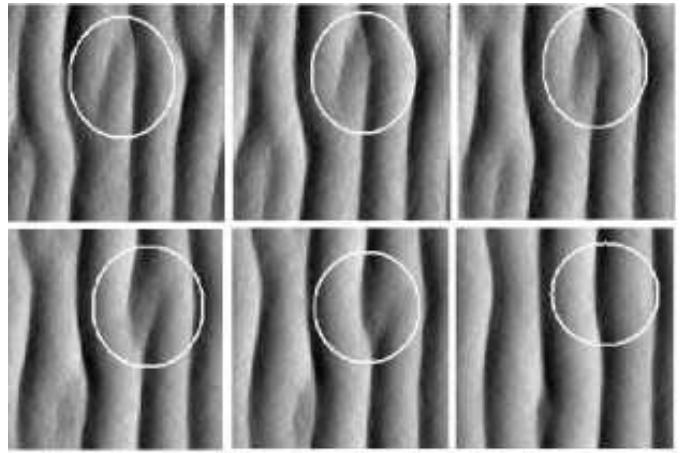}}
\vskip 0.2true cm
\caption{The disappearance of a kink. {\it (top row)}
The smallest branch of the ripple climbs on the main part.
{\it (bottom row)} the avalanche feeds the ripple, and the branch vanishes.}\label{k1}
\end{center}
\end{figure}

The disappearance of a defect mainly occurs when a ripple is splitted into two inequitably parts (see an example in Figure \ref{k1}). The small branch moves faster than the larger one and climbs on it following the merging dynamics described in section IV.A. Reaching the crest, the small branch creates an avalanche. The sand rolling on the lee slope may be captured. If the amount of sand escaped from this avalanche is small, the defect disappears and the small branch vanishes. After this process, the number of defects is reduced. If all defects show this kind of behavior the defect density will rapidly tend to zero. However, this is not the case \cite{caps}.

\begin{figure}[H]
\begin{center}
\centerline{\epsfxsize=9.cm \epsfysize=6.cm
\epsffile{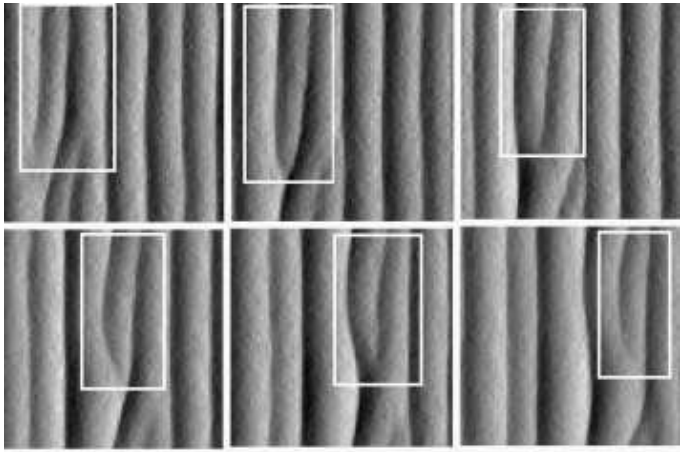}}
\vskip 0.2true cm
\caption{The propagation of a kink. The defect is expelled downstream.}\label{k2}
\end{center}
\end{figure}

The propagation of a kink takes place when a sufficient amount of sand escapes from the avalanche. This is the case when the branches are quite long. In the emphasized part of Figure \ref{k2}, one can see the ending of a small ripple. The extremity of the ripple moves faster than the main part. The merge (second image) is accompanied by an avalanche. As seen in the previous paragraph, sand is expelled out of the avalanche. But here, the amount of sand is sufficient to create a new ripple. The merging/propagation process is thus repeated on the next ripple.

Let us consider a branch of length $\ell_b$. The merge of this branch with a larger ripple causes the decrease of the branch length since a part of the branch has been owned by the ripple. When a second meeting occurs, $\ell_b$ is once again reduced. As a consequence, the initial length of the branch is a relevant parameter in order to predict the life-time of a defect. For example, a kink of initial length $\ell_b=65$ can typically propagate during $9.5\times 10^5$ iterations, while a kink of initial length $\ell_b=23$ has a life expectancy of $2.5\times 10^5$ iterations. One should note that those life-times mainly depend on the diffusion coefficient $D$ and the avalanche process.

In Ref.\cite{caps}, we have found a non zero asymptotic value for the defect density. The existence of propagating kinks gives a picture of this behavior. The relevant ingredient for kink disappearance being the length of the branch.  

\section{Summary}

We have introduced new ingredients in the Nishimori-Ouchi (NO) model:  the existence of an angle of repose $\theta_r$ and subsequent avalanches when the local slope $\theta$ is larger than $\theta_r$. The new model reproduces realistic features of aeolian ripples such as a non-linear evolution of the ripple amplitude $h_{max}$. The origin of this behavior has been explained by the merge of ripples traveling at different speeds. Increasing the saltation length, we have observed a grows of the ripple wavelength, in agreement with observations. In the case of a constant wind orientation, natural ripples are asymmetric. This feature has also been reproduced by the new model. 

Studying the kink dynamics, we have shown the coexistence of two kinds of kink evolution: propagation and disappearance. The defect initial length has been proposed as a relevant parameter in order to predict the life-time of that defect. From this result, the existence of a non-zero asymptotic value of the kink density has been explained.

\section*{Acknowledgements}

HC is financially supported by the FRIA, Belgium. This work is also
supported by the Belgian Royal Academy of Sciences through the
Ochs-Lefebvre prize.

\end{document}